\documentclass[12pt]{article}
\pdfoutput=1

\usepackage[pdftex]{graphicx}
\usepackage{lscape} 
\usepackage{array}

\usepackage{fix-cm}

\usepackage{caption}
\usepackage[makeroom]{cancel}
\usepackage[normalem]{ulem}
\usepackage{amsmath,amssymb}
\usepackage{slashed}
\usepackage{nicefrac, xfrac}
\usepackage{xcolor}
\usepackage{cite}
\usepackage{pdflscape}
\usepackage{multirow}
\usepackage[thinlines]{easytable}
\usepackage{url}
\usepackage[utf8]{inputenc}
\usepackage{longtable}
\usepackage{booktabs}
\usepackage[hidelinks]{hyperref} 

\newcommand{\eq}[1]{\begin{equation} #1 \end{equation}}

\newcommand{\Fig}[1]{Figure~\ref{#1}}
\newcommand{\cL}{{\cal L}}

\makeatletter
\g@addto@macro\bfseries{\boldmath}
\makeatother

\begin{document}


\begin{titlepage}

\vspace*{-2cm}
\begin{flushright}
\end{flushright}

\vspace{2.2cm}

\begin{center}
\bf
\fontsize{19.6}{24}\selectfont
Three-loop contributions to $b\to s\gamma$ associated with the current-current operators
\end{center}

\vspace{0cm}

\begin{center}
\renewcommand{\thefootnote}{\fnsymbol{footnote}}
             {Christoph Greub$^a$, Hrachia M. Asatrian$^b$, Hrachya H. Asatryan$^{c}$, \newline
               Lukas Born$^a$ and Julian Eicher$^a$}
\renewcommand{\thefootnote}{\arabic{footnote}}
\setcounter{footnote}{0}

\vspace{.8cm}
\centerline{${}^a$\it Albert Einstein Center for Fundamental Physics, Institute for Theoretical Physics,}
\centerline{\it University of Bern, CH-3012 Bern, Switzerland}
\vspace*{2.5mm}
\centerline{${}^b$\it Yerevan Physics Institute, 0036 Yerevan, Armenia}
\vspace{2.5mm}
\centerline{${}^c$\it Yerevan State University, 0070 Yerevan, Armenia}
\vspace{2.5mm}

\vspace*{.2cm}

\end{center}

\vspace*{10mm}
\begin{abstract}\noindent\normalsize
  In a recent work, we calculated all three-loop diagrams contributing to
  the decay amplitude for $b \to s \gamma$ where none of the gluons touch the $b$-leg.
  In the present paper, we complete the calculation by working out all remaining three-loop diagrams (of order $\alpha_s^2$)
  associated with the current-current operators $O_1$ and $O_2$ at the physical value of the charm-quark mass $m_c$.
  Using the programs \texttt{AMFlow} and \texttt{DiffExp} to solve the differential equations for the master integrals,
  we obtained precise numerical results at 23 values for $z=m_c^2/m_b^2$, ranging from $z=1/1000$ to $z=1/5$, along with asymptotic expansions around $z=0$.
  For certain diagrams, the asymptotic expansion breaks down in the physical $z$-range, necessitating a Taylor expansion (which we do around $z=1/10$).
  In all expansions, we retained power terms up to $z^{20}$ and included the accompanying $\log(z)$ terms to all powers
  for asymptotic expansions. Numerical results for the sum of all diagrams (including those calculated in the previous paper) are presented in tabular form,
  while the mentioned expansions of individual diagram classes are provided electronically. We note that our results for the asymptotic expansions around $z=0$
  are in good agreement with those recently published by Fael {\it et al.} and Czaja {\it et al.}.

\end{abstract}

\end{titlepage}

\renewcommand{\theequation}{\arabic{section}.\arabic{equation}} 

\setcounter{tocdepth}{2}


\section{Introduction}
\label{sec:intro}
\setcounter{equation}{0}
Rare $B$-meson decays have been the focus point of theorists and experimentalists for some time, due to the potential to test the
Standard Model (SM) at scales of several hundreds of GeV. The decay $B\to X_s \gamma$ is particularly suitable as it allows a very stringent comparison of 
experimental and theoretical information. Both experiment and theory have made substantial progress over the last two decades and have already reached an
impressive precision which puts strong constraints on extensions of the SM. In view of further increasing precision for the experimental measurements, 
refined theoretical predictions are required. This is possible because the (inclusive) decay $B\to X_s \gamma$ can be approximated by the quark-level process $b \to X_s \gamma$
which can be treated by perturbative QCD in the effective theory obtained after integrating out the heavy particles $t, W, Z$ and $H$. 
 
At the present or expected precision of the experimental measurements of
the decay $B\to X_s \gamma$, a full next-to-next-to-leading logarithmic order (NNLL) calculation
is necessary to reduce the theoretical uncertainties and to make a rigorous
comparison with existing and future experimental data. 

A first estimate of the branching
ratio at NNLL level, leading to ${\cal B}(B\to X_s \gamma)=(3.15\pm 0.23)\times 10^{-4}$,
was done in \cite{Misiak:2006zs}. An updated version for this
branching ratio, incorporating those results for NNLL contributions and lower-order perturbative corrections that had been calculated after 2006,
was provided in \cite{Misiak:2015xwa} where some of us were involved. The CP- and isospin-averaged branching ratio reads
${\cal B}(B\to X_s \gamma)=(3.36\pm 0.23)\times 10^{-4}$ which is in agreement with the current experimental
average ${\cal B}(B\to X_s \gamma)=(3.49\pm 0.19)\times 10^{-4}$
given in \cite{HFLAV}. We note that 
all of these numbers refer to a cut of the photon energy below $1.6$ GeV.

A sizeable part of the uncertainty in \cite{Misiak:2015xwa} is due to the fact that so far there is no exact calculation with the correct mass for the charm quark $m_c$ at NNLL
order. Instead, the results are obtained via interpolation, i.e., by using the results obtained through the large $m_c$ asymptotic expansion on one hand 
and the results for $m_c=0$ on the other hand. An improvement resulted from calculating a limited set of diagrams with closed fermion loops on gluon lines
at order  $\alpha_s^2$ for the physical value of $m_c$ in \cite{Misiak:2020vlo}\footnote{The diagrams calculated are those occurring in the 
interference of the current-current and photonic dipole operators.}; see also \cite{Ligeti:1999, Bieri:2003ue, Boughezal:2007}.
Let us remind the reader that the complete
NNLL calculation of the branching ratio includes a large number of diagrams associated with different effective operators,
 which pose their own difficulties that have to be overcome. 

We feel that the time has come to close the gap on $m_c$. In \cite{Greub:2023msv} we started the computation of virtual $\alpha_s^2$ corrections
to the decay amplitude $b \to s  \gamma$  associated with the current-current operators, that is the matrix elements of these operators. 
We worked out all those diagrams where no gluons are touching the $b$-quark line.
These contributions, which we denote in the following as ``$s$-leg diagrams'',
are shown in Figure 1 of \cite{Greub:2023msv}. We also set out to calculate the remaining diagrams\footnote{This work can be considered as an extension of our work \cite{Greub:1996tg}, where the corresponding  $\alpha_s^1$ corrections were calculated.}.

A few months later, two papers were published \cite{Fael:2023,Czaja:2023ren} in which our results were confirmed and extended to the complete set of virtual correction diagrams.
As these highly non-trivial computations are ingredients in the NNLL program which need checking, we decided to continue an independent calculation of the
contributions which were not included in our paper \cite{Greub:2023msv}.
To this end, we evaluate in the present paper all diagrams of order $\alpha_s^2$ in \Fig{fig:diags_bleg}
where no gluon touches the $s$-quark line (referred to as ``$b$-leg diagrams''), as well as the ``mixed diagrams'' in \Fig{fig:diags_mixed}, and finally
the ``bubble diagrams'' in \Fig{fig:diags_bubble} where the gluon propagator is dressed by a fermion-, a gluon- or a ghost-loop.

The remainder of this paper is organized as follows: In Section 2 we present the theoretical framework and a few conventions. In Section 3 we present the
methodology of solving differential equations using the program \texttt{AMFlow}, as well as checks we have done using the program \texttt{DiffExp}. In Section 4, we present
numerical results for $z \doteq m_c^2/m_b^2=0$ and at 23 points between $z=1/1000$ and $z=1/5$. We also provide
the result as an asymptotic expansion around $z=0$ (which contains non-negative integer and half-integer powers of $z$ as well as non-negative integer powers of $\log(z)$).
As it turns out that this expansion
breaks down in the physical $z-$range
for some ``$b$-leg'' diagrams, we also provide a Taylor expansion for this class around the value $z=1/10$
(which is a typical physical value for $z$). We illustrate with a plot that the transition between the two mentioned formulas is very smooth.
We note that the authors of \cite{Fael:2023} also worked out several expansions; they explicitly gave the result for the asymptotic expansion around $z=0$,
which is in perfect agreement with our formula.
In Section 5 we summarize our work. The results for different classes of diagrams are submitted in electronic form
together with the paper, as described in the Appendix.

\section{Theoretical framework}
\label{sec:framework}
\setcounter{equation}{0}

$B$-meson or $b$-quark decay amplitudes are usually calculated within the Weak Effective Theory (WET) where the SM particles with EW-scale masses
have been integrated out. The WET Lagrangian then contains QCD and QED interactions, and a tower of higher dimensional local operators which is
typically truncated at dimension six~\cite{Buchalla:1995vs,Aebischer:2017gaw}. The part of the WET Lagrangian which is relevant for the
contributions discussed in this paper is
\eq{
\cL_{\rm WET} = \cL^{(4)} + \frac{4G_F}{\sqrt2} V_{ts}^*V_{tb} \Big[
C_1 O_1 + C_2 O_2  + C_7 O_7 
\Big] \, ,
\label{LWET}
}
where
\begin{align}
&O_1 = (\bar s \gamma_\mu P_L T^a c)(\bar c \gamma^\mu P_L T^a b)\ ,
&&O_2 = (\bar s \gamma_\mu P_L c)(\bar c \gamma^\mu P_L b) \ ,
\nonumber\\[2mm]
&O_7 = \frac{e}{16\pi^2}m_b(\mu)(\bar s \sigma_{\mu\nu} P_R b) F^{\mu\nu}\ .
&&
\end{align}
$\cL^{(4)}$ contains the usual kinetic terms and the mass terms of the quarks $u, d, s, c, b$ as well
as their interactions with the photon and the gluons.
$P_{R,L}=(1\pm\gamma_5)/2$ stand for the right and left projection operator, $\sigma_{\mu\nu}\equiv (i/2) [\gamma_\mu,\gamma_\nu]$ and our convention for the covariant
derivative is given by $D_\mu q = (\partial_\mu + i e Q_q A_\mu + i g_s T^A G^A_\mu)q$; $m_b(\mu)$ in the definition of $O_7$ denotes the mass of the $b$-quark
in the $\overline {\rm MS}$-scheme. Note that what is calculated in this paper are only the bare (i.e. unrenormalized) $\alpha_s^2$ corrections from $O_{1,2}$ to the decay amplitude for $b \to s \gamma$. In these terms the renormalization scheme of $m_b$ and $m_c$ is not fixed (scheme differences would result in effects of $O(\alpha_s^3)$ ).
Furthermore, we will neglect the strange quark mass throughout our paper.

\subsection{Form factor decomposition}
\label{subsec:formfactor}
As mentioned in some detail in \cite{Greub:2023msv}, the decay
amplitude ${\cal A}(b \to s \gamma)=\langle s \gamma|O_{1,2}|b\rangle$ can be written as ${\cal A}=M_{\mu} \, \varepsilon^\mu$, where  $\varepsilon^\mu$
denotes the polarization vector of the emitted photon. In the considered limit where the strange quark mass is put to zero,
$M_{\mu}$ is of the form
\eq{
M_{\mu} =  \bar u_s (p_s) P_R  \left[ A \, q_\mu + B \, p_\mu + C \, \gamma_\mu \right] u_b(p)\ . \label{Mmui}
}
In this equation $u_b(p)$ denotes the Dirac spinor of the $b$-quark with four-momentum $p$, $u_s(p_s)$ is the analogous quantity
for the $s$-quark with four-momentum $p_s$ and $q=p-p_s$ is the four-momentum of the emitted photon. Applying standard algebraic manipulations, the form factors  $A$, $B$, $C$
are given in terms of linear combinations of scalar three-loop integrals\footnote{The form of such integrals can be seen e.g. in Eq. (3.4) in our previous paper \cite{Greub:2023msv}.}. After performing these integrals and taking into
account that $q^2=0$ in our process, these form factors are functions of $m_b$ and $m_c$ (and depend also on the renormalization scale $\mu$).
The function $A$ does not contribute to $b \to s \gamma$ since the corresponding tensor structure vanishes when contracting $M_{\mu}$ with
  the polarization vector. Furthermore, the Ward identity $q^{\mu} M_{\mu}=0$ (resulting from electromagnetic gauge invariance) implies that
\eq{C=-\frac{m_b}{2} \, B \, .\label{Wardrelation} }
This Ward identity for on-shell amplitudes, derived in detail for instance in sections 7.4 and 10 of the text-book by Peskin and Schroeder \cite{Peskin}, holds
for renormalized amplitudes in which the masses of external particles are defined in the on-shell scheme.
In our present paper, we only calculate the bare version of the form factor $B$. When later including the renormalization of the amplitude,
we suggest to calculate the effects of the counterterms on $B$ only and then {\it use} the relation (\ref{Wardrelation}) to get the renormalized version of $C$.
This procedure \`a priori leads to expressions for renormalized form factors $B$ and $C$ in the on-shell scheme for $m_b$. By performing a further appropriate finite
renormalization, one can transform the form factors to any other renormalization scheme for $m_b$.

Two remarks concerning the diagrams leading to the bare version of the form factor $B$ are in order, which
are best illustrated in Figure 1 of \cite{Greub:2023msv}: 
first, it is quite easy to see that those diagrams which are marked with a cross that does not
carry a number would only contribute to the form factor $C$. Therefore only the diagrams with numbered crosses (1-44) had to be
worked out. Second, diagrams which contain a 1-loop quark self-energy on the very left or very right of the external $b$- or $s$-leg (i.e. which formally
contain a zero propagator denominator)
are omitted from the list of bare diagrams; their effects can be taken into account later by quark wave function renormalization. 
Note that in the present paper we only listed those (numbered) diagrams which contribute to the bare form factor $B$. For the calculation of this form factor we
employ the method of differential equations, which is detailed in Section \ref{sec:DiffEqs}.

\begin{figure}
\begin{center}
\includegraphics[width=5cm]{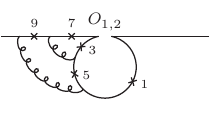} \quad
\includegraphics[width=5cm]{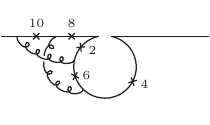} \quad
\includegraphics[width=5cm]{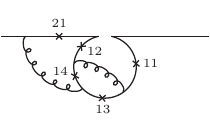}
\end{center}
\begin{center}
\includegraphics[width=5cm]{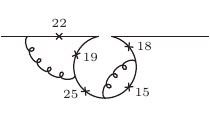} \quad
\includegraphics[width=5cm]{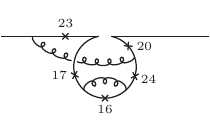} \quad
\includegraphics[width=5cm]{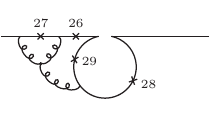}
\end{center}
\begin{center}
\includegraphics[width=5cm]{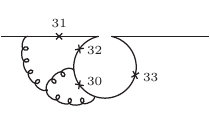} \quad
\includegraphics[width=5cm]{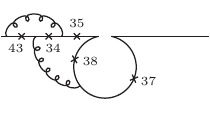} \quad
\includegraphics[width=5cm]{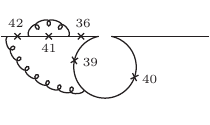}
\end{center}
\begin{center}
\includegraphics[width=5cm]{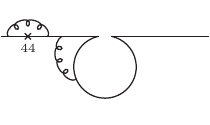}
\end{center}
\caption{``$b$-leg diagrams'': List of three-loop contributions to $b \to s \gamma$ associated with $O_1$ and $O_2$, where no gluon
  touches the $s$-leg. A cross on a quark line represents a possible place where the photon can be emitted. Only the diagrams which contribute to the form factor $B$ are shown; they carry a diagram number next to the cross. See also the text at the end of Section \ref{subsec:formfactor}.}
\label{fig:diags_bleg}
\end{figure}

\begin{figure}
\begin{center}
\includegraphics[width=5cm]{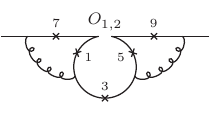} \quad
\includegraphics[width=5cm]{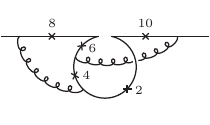} \quad
\includegraphics[width=5cm]{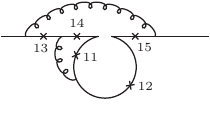}
\end{center}
\begin{center}
\includegraphics[width=5cm]{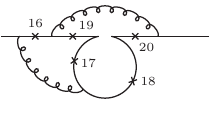} \quad
\includegraphics[width=5cm]{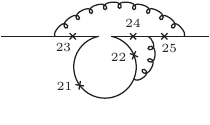} \quad
\includegraphics[width=5cm]{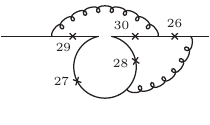}
\end{center}
\begin{center}
\includegraphics[width=5cm]{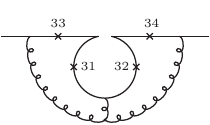} \quad
\includegraphics[width=5cm]{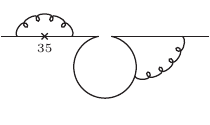} \quad
\includegraphics[width=5cm]{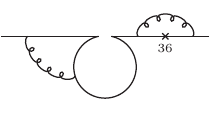}
\end{center}
\caption{``Mixed diagrams'': List of three-loop contributions to $b \to s \gamma$ associated with $O_1$ and $O_2$, where one gluon
  touches the $b$-leg and one touches the $s$-leg. A cross on a quark line represents a possible place where the photon can be emitted. Only the diagrams which contribute to the form factor $B$ are shown.}
\label{fig:diags_mixed}
\end{figure}

\begin{figure}
\begin{center}
\includegraphics[width=5cm]{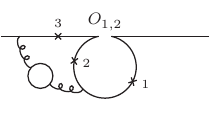} \quad
\includegraphics[width=5cm]{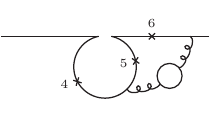}
\end{center}
\caption{List of the ``bubble diagrams'' contributing to $b \to s \gamma$ associated with $O_1$ and $O_2$.
  A cross on a quark line represents a possible place where the photon can be emitted. The particle running in the bubble (inserted in the gluon line) is one of the following:
  a massless fermion ($u,d,s$-quarks), a fermion with mass $m_c$, a fermion with mass $m_b$, a gluon, or a ghost. Only the diagrams which contribute to the form factor $B$ are shown.}
\label{fig:diags_bubble}
\end{figure}

\subsection{Comment on the computation of bubble diagrams}

We note that for the
``bubble diagrams'' in \Fig{fig:diags_bubble} it is sufficient to calculate only the diagrams where the gluon propagator is dressed by a fermion loop; as has been shown in
\cite{Blokland:2005uk, Asatrian:2006sm, Muta:1987},
the sum of the ghost- and the gluon-loop can be obtained from the massless fermion loop diagrams by replacing (in the Feynman gauge)
\begin{equation}
     n_\ell \, tr \, \to -C_A \, \left( \frac{5}{4}+\frac{\epsilon}{2}+\frac{\epsilon^2}{2}+\frac{\epsilon^3}{2}+O(\epsilon^4) \right) \, ,
\end{equation}
where $n_\ell$ denotes the number of massless quarks and $C_A$ and $tr$ are color factors.

\section{Differential equations}
\label{sec:DiffEqs}
\setcounter{equation}{0}

In this section, we explain the details of the calculations of the form factor $B$ using the method of differential equations. The general outline of this method is as follows:
express each diagram (or sets of diagrams) as a linear combination of scalar integrals,
which in turn are reduced to a smaller number of master integrals (MIs) by applying Integration-By-Parts (IBP) identities to the scalar integrals
(see \cite{Klappert:2020nbg, Maierhofer:2017gsa, Lewis, Klappert:2020aqs, Klappert:2019emp, Usovitsch, Laporta:2000dsw, Lee:2012cn, Lee:2013mka}).
We use the program \texttt{Kira} \cite{Maierhofer:2017gsa} to bring the scalar integrals to a set of MIs.
In general, there are many different sets (we can call each set a basis) of MIs we can reduce to. In \cite{Usovitsch},
an algorithm is suggested which chooses a ``good'' basis in the sense that the $\epsilon$ dependence
factors out in the denominators of the relations in which the scalar integrals are expressed in terms of MIs;
this simplifies further steps in the calculation, in particular solving the differential equations. We therefore
use this ``good'' basis throughout our calculations.

Using the IBP identities generated by \texttt{Kira}, one can construct a system of differential equations for the MIs
with respect to $z$. These equations are of the form
\begin{equation}
    \partial_z J_k(\epsilon, z) = a^{k\ell}(\epsilon, z) J_{\ell} (\epsilon, z) ,
\label{eq:derivative}
\end{equation}
where $a^{k\ell}$ are the entries of a $N\times N$ matrix depending on $\epsilon$ and $z$. The derivatives of the MIs $J_k$ are obtained by differentiating
the integrands, which produce new scalar integrals. These scalar integrals are then subjected to IBP reduction again to express the
derivatives $\partial_z J_k$ in terms of the MIs $J_\ell$ (for more detail, see \cite{Greub:2023msv}). 

In \cite{Greub:2023msv}, two methods were used for solving the differential equations. For most diagrams, we were able to transform the differential
equation into canonical form \cite{Meyer:2017joq}, where the equations could then be solved in an iterative manner as an expansion in $\epsilon$.
The $z$ dependence in these solutions are contained in Generalized Polylogarithms (GPLs) \cite{Goncharov:1998kja}. After fixing the integration
constants (using the large $m_c$ behavior of the MIs), we were left with purely analytical precise results.
However, for two sets of diagrams (namely (11, 12) and (13, 14)), we were unable to transform the differential equations to canonical form. As such,
a different method was used to solve these differential equations: a series expansion was constructed around $z=0$, by bringing the differential
equation matrix first to Fuchsian form \cite{Lee:2020zfb} and then to Jordan form. This leads to a simple first-order linear differential equation,
which has $N$ linearly independent fundamental solutions. To fix the integration constants, the program \texttt{FIESTA5} \cite{Smirnov:2021rhf} was used. This program
allows to numerically calculate the leading terms of the $z$-expansion of the MIs directly from their integral representations. This entire process is
discussed in detail in \cite{Greub:2023msv}. The problem with this second approach is the precision of \texttt{FIESTA5}; the authors of \cite{Fael:2023} state that when
comparing with the results of our ``$s$-leg diagrams'', for the sets of diagrams involving \texttt{FIESTA5} results, the agreement was only 5 digits,
while for other sets, it was at least 10 digits. 
Given the extra complexity of the classes of diagrams we solve in the present paper, it is problematic to use the first method (that is, solving the
differential equations analytically) for most sets of diagrams, forcing us to use the second method which involves \texttt{FIESTA5}.
We found, however, that the results suffer from very low precision; therefore we use a different
method for the solution of our differential equations, which is discussed in the following subsections.

\subsection{Calculation of boundary conditions for differential equations}
\label{sec:BoundaryConditions}
To solve the system of differential equations for a given diagram in the three classes considered in this paper, we need $N$ boundary conditions, where
$N$ is the number of MIs in the diagram. We use
the program \texttt{AMFlow} \cite{Liu:2022chg}, which is based on the methods developed in \cite{Liu:2017jxz}, to calculate these boundary conditions at a specific point.
In our application
\texttt{AMFlow}
receives as input the value of $m_c^2$ and $m_b^2$ (or the value of $z=m_c^2/m_b^2$, when working in units where $m_b$ is put to $1$, as we do),
and calculates the given list of master integrals for these values of the quark masses with very high
precision as a Laurent series of sufficiently high order in $\epsilon$. 

In practice, we used the \texttt{AMFlow} package to calculate the boundary values of the MIs at $z=1/100$.

\subsection{Solving the differential equations}
\label{sec:diffeqsols}
To solve the differential equations for a given diagram in the three classes considered in this paper, we use the \texttt{DESolver} package of \texttt{AMFlow}. \texttt{DESolver} allows us to
transport the results for the MIs from $z=1/100$ to any point. We use this functionality to iteratively obtain the results for the MIs at the following 23 points:
\begin{align}
\biggl\{ \frac{1}{1000}, \frac{1}{500}, \frac{1}{200}, \frac{1}{100}, \frac{2}{100}, \frac{3}{100}, \frac{4}{100}, \frac{5}{100}, \frac{6}{100}, \frac{7}{100}, \frac{8}{100}, \frac{9}{100},
\nonumber\\
\frac{10}{100}, \frac{11}{100}, \frac{12}{100}, \frac{13}{100}, \frac{14}{100}, \frac{15}{100}, \frac{16}{100}, \frac{17}{100}, \frac{18}{100}, \frac{19}{100}, \frac{20}{100} \biggr\}
\label{eq:points}
\end{align}
To validate these results, we also directly calculated the values of the MIs at $z=1/10$ using \texttt{AMFlow}. When comparing the directly calculated
values at $z=1/10$ with those obtained via transportation from $z=1/100$, we have perfect agreement; this consistency makes us confident that
the results for the MIs for each diagram at the 23 points mentioned above are very precise. This in turn also means that we have very precise numerical results for
the contribution of the three classes of diagrams to the form factor $B$ (see Eq. (\ref{Mmui})) at these 23 points.

\begin{figure}
\begin{center}
\includegraphics[width=8cm]{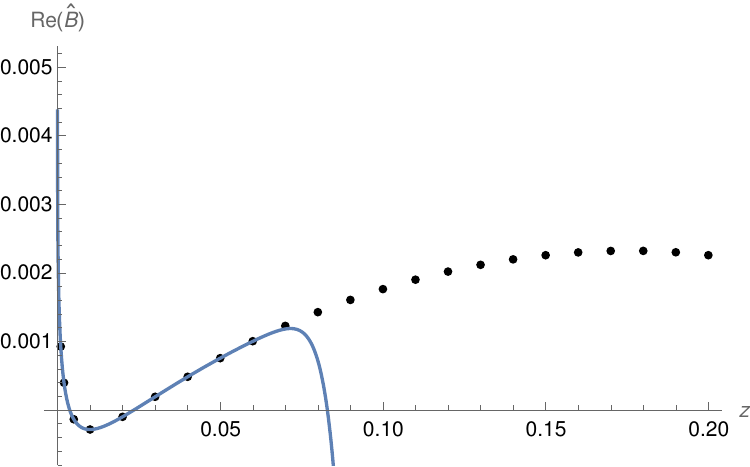}
\end{center}
\caption{Real part of the $\epsilon^0$ coefficient of the $O_2$ contribution to $\hat{B}$ as defined in Eq. (\ref{eq:Bhat}). The highlighted points are the 23 points mentioned in the text.
  The solid line is the sum of the expansion around $z=0$ for all four classes of diagrams. As can be seen, the asymptotic expansion breaks down
  around the value of $z=0.07$.}
\label{fig:res_asym}
\end{figure}

\begin{figure}
\begin{center}
\includegraphics[width=8cm]{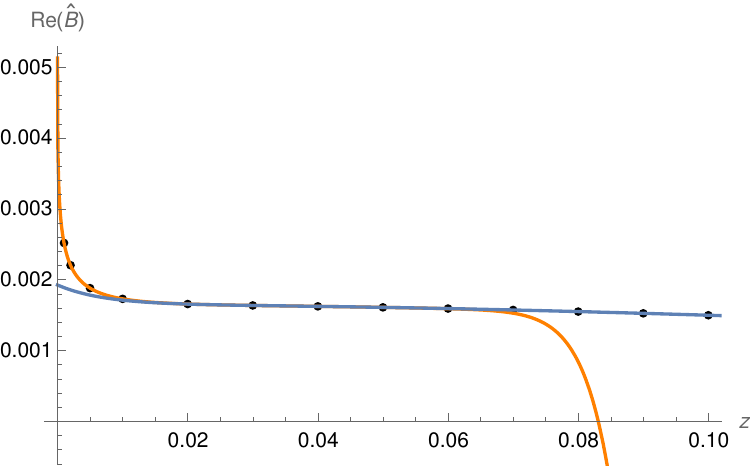}
\end{center}
\caption{Real part of the $\epsilon^0$ coefficient of the $O_2$ contribution to $\hat{B}$ as defined in Eq. (\ref{eq:Bhat}) for all ``$b$-leg diagrams''. The highlighted points are the
  23 points mentioned in the text. The orange line shows the sum of the asymptotic expansion for all diagrams in Figure \ref{fig:diags_bleg}, while the blue
  line shows the corresponding Taylor expansion around $z=1/10$. The breakdown of the asymptotic expansion near the physical range of $z$, as well as the breakdown
  of the Taylor expansion near $z=0$ are clearly visible.
  The transition from the asymptotic expansion to the Taylor expansion is very smooth.}
\label{fig:res_taylor}
\end{figure}

Besides these numerical results, it is convenient to have also an analytic expression in terms of an asymptotic expansion around $z=0$ for these contributions to the form factor $B$.
In particular, such an expression also covers the results for the (CKM unsuppressed) $u$-quark loops in the analogous process $b \to d \gamma$, where the corresponding $z$ is $m_u^2/m_c^2$, which
is basically zero. The asymptotic expansion around $z=0$
can easily be constructed on a diagram-by-diagram basis with the program \texttt{DESolver}, using as input the differential equation matrix,
as well as the boundary conditions for the MIs which we have calculated at $z=1/100$.
We worked out this asymptotic expansion for the contribution to the form factor $B$ of all three classes of diagrams.
For the ``mixed diagrams'' and for the ``bubble diagrams'', the obtained asymptotic formulas coincide very well with the 23 points.
Only for ``$b$-leg diagrams'', we have noticed that this expansion breaks
down within the physical $z$-range\footnote{We use numerical values $m_c \sim 1.0 \rm{ \ GeV},...,1.7 \rm{ \ GeV}$ and $m_b \sim 4.2 \rm{ \ GeV},...,5.0 \rm{ \ GeV}$ which correspond to $z \in [0.04,0.16]$.}.
Detailed investigations show that the convergence radius is not large enough
to cover the full relevant $z$-range for 
some of the master integrals associated with the diagrams 3-6; this is   
related to the presence of poles at $z=1/16$ in
certain entries of the matrices which define the differential equations.
In Figure \ref{fig:res_asym}, we combine the $\epsilon^0$ coefficients of the three classes of diagrams with the ``$s$-leg diagrams'' computed in \cite{Greub:2023msv}; the
breakdown of the asymptotic expansion is clearly seen. 
Given that the physical value for $z$ is $\sim 1/10$ (i.e. higher than the value for which the $z=0$ expansion breaks down), we need a different expansion.
To this end, we also worked out a Taylor expansion for the ``$b$-leg diagrams'' around $z=1/10$, which is a regular point of the differential equations; this expansion is also implemented in the \texttt{DESolver}
package of \texttt{AMFlow} and uses as input the values of the MIs at the expansion point. Alternatively, this expansion could also be constructed ``by hand'':
Given the input of the MIs at $z=1/10$, their first derivatives with respect to $z$ can be obtained at $z=1/10$ from the right-hand side of eq. (\ref{eq:derivative}).
By interating the differential equation, one also finds the second derivatives at $z=1/10$, the third derivatives and so on. Having these derivatives at hand,
it is straightforward to set up the Taylor expansion.
In Figure \ref{fig:res_taylor}, we illustrate that the transition between these two formulas is very smooth. We therefore propose to give the final results
for the ``$b$-leg diagrams'' in terms of a ``switched expansion'', by switching at $z=0.04$ from the asymptotic expansion to the Taylor expansion,
as suggested by the figure.
Using these two expansions, we are able to
cover all values of $z$ from $0$ up to and beyond the physical range. The two expansions are given up to order $z^{20}$ and $(z-1/10)^{20}$, respectively, after having checked that this accuracy is sufficient. For the asymptotic expansion we have kept the accompanying $\log(z)$ terms to all powers of $z$.
All our results are presented in the electronic file submitted with the paper.

As mentioned at the beginning of this section, the results for ``$s$-leg diagrams'' 11-14 in \cite{Greub:2023msv} are not very accurate due to limitations
of the \texttt{FIESTA5} program. Therefore, we have recalculated these four diagrams using our new method. These new calculations are very precise
and were used in the final results given in the electronic file submitted with the paper.

Note that we have calculated most of the diagrams also using the program \texttt{DiffExp} \cite{Hidding:2020ytt}.

\section{Results}
\label{sec:Result}
\begin{figure}
\begin{center}
\includegraphics[width=8cm]{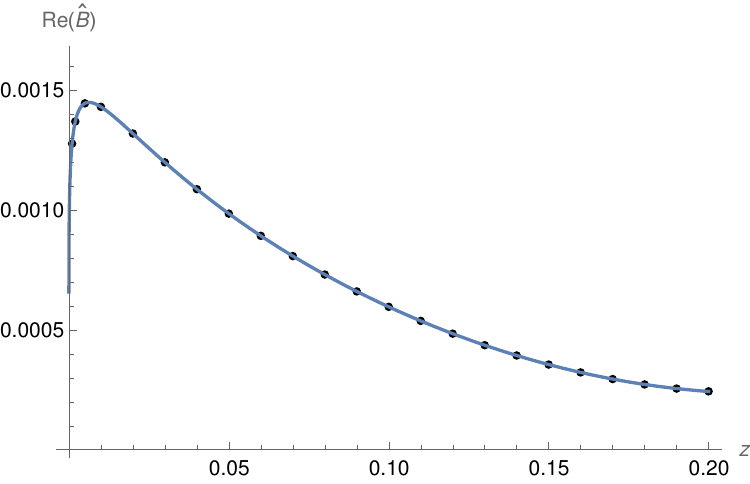}
\includegraphics[width=8cm]{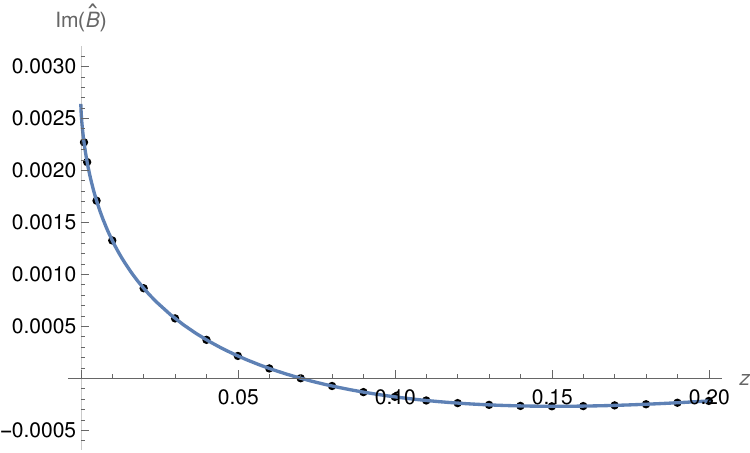}
\end{center}
\caption{$\epsilon^0$ coefficient of the $O_1$ contributions to $\hat{B}$ as defined in Eq. (\ref{eq:Bhat}). The highlighted 23 points are mentioned in the text. The solid line is
  constructed by switching at $z=0.04$ from the asymptotic expansion to the Taylor expansion.}
\label{fig:res_o1}
\end{figure}

\begin{figure}
\begin{center}
\includegraphics[width=8cm]{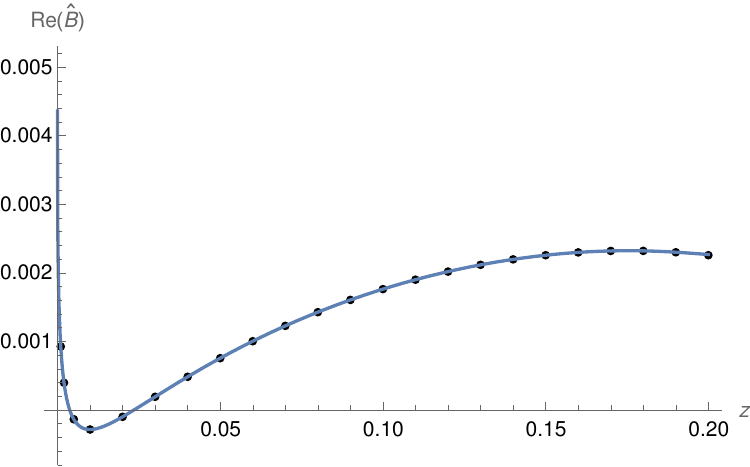}
\includegraphics[width=8cm]{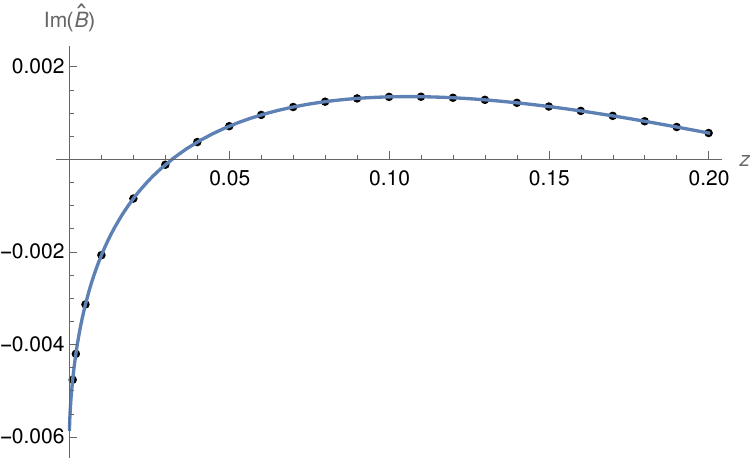}
\end{center}
\caption{$\epsilon^0$ coefficient of the $O_2$ contributions to $\hat{B}$ as defined in Eq. (\ref{eq:Bhat}). The highlighted 23 points are mentioned in the text.
  The solid line is constructed by switching at $z=0.04$ from the asymptotic expansion to the Taylor expansion.}
\label{fig:res_o2}
\end{figure}

\setcounter{equation}{0}
In this section we present the contributions of all four classes of diagrams of order $\alpha_s^2$ to the form factor $B$, both for $O_1$ and $O_2$.
In fact, we show our results in terms of
the dimensionless quantity $\hat{B}$, defined according to
\eq{B = {\rm pref} \cdot \hat{B} \quad \mbox{with} \quad {\rm pref}=-\frac{e \, m_b}{4 \pi^2} \, g_s^4\,\left( \frac{\mu^2}{m_b^2} \right)^{3 \epsilon} \, . \label{eq:Bhat}}
In Tables \ref{tab:table_O1} and \ref{tab:table_O2} the dimensionless form factor $\hat{B}$ is given (for all powers of $\epsilon$ up to $\epsilon^0$) at the 23 points
mentioned in Eq. (\ref{eq:points}) for the operators $O_1$ and $O_2$, respectively.
\begin{table}[]
    \centering
    \begin{tabular}{|c|c|c|c|}
    \hline
     $O_1$ &  $\frac{1}{\epsilon^{2}}$ & $\frac{1}{\epsilon^{1}}$ &  $\epsilon^{0}$  \\ [1ex] 
     \hline\hline
         $\frac{1}{1000}$ & $ 0.02522-0.02553 i $ & $ 0.3025+0.08587 i $ & $ 1.278+2.269 i $ \\
    \hline
         $\frac{1}{500}$ & $ 0.02572-0.02425 i $ & $ 0.3025+0.07716 i $ & $ 1.371+2.079 i $ \\
    \hline
         $\frac{1}{200}$ & $ 0.02716-0.02153 i $ & $ 0.2936+0.05769 i $ & $ 1.447+1.707 i $ \\
    \hline
         $\frac{1}{100}$ & $ 0.02935-0.01845 i $ & $ 0.2732+0.03502 i $ & $ 1.432+1.325 i $ \\
    \hline
         $\frac{2}{100}$ & $ 0.03314-0.01437 i $ & $ 0.2322+0.005243 i $ & $ 1.321+0.8652 i $ \\
    \hline
         $\frac{3}{100}$ & $ 0.03640-0.01156 i $ & $ 0.1946-0.01436 i $ & $ 1.200+0.5751 i $ \\
    \hline
         $\frac{4}{100}$ & $ 0.03931-0.009440 i $ & $ 0.1602-0.02823 i $ & $ 1.088+0.3689 i $ \\
    \hline
         $\frac{5}{100}$ & $ 0.04194-0.007765 i $ & $ 0.1287-0.03832 i $ & $ 0.9862+0.2139 i $ \\
    \hline
         $\frac{6}{100}$ & $ 0.04435-0.006405 i $ & $ 0.09965-0.04568 i $ & $ 0.8934+0.09385 i $ \\
    \hline
         $\frac{7}{100}$ & $ 0.04659-0.005284 i $ & $ 0.07275-0.05097 i $ & $ 0.8089-0.0005617 i $ \\
    \hline
         $\frac{8}{100}$ & $ 0.04867-0.004348 i $ & $ 0.04775-0.05465 i $ & $ 0.7317-0.07520 i $ \\
    \hline
         $\frac{9}{100}$ & $ 0.05062-0.003563 i $ & $ 0.02447-0.05704 i $ & $ 0.6612-0.1340 i $ \\
    \hline
         $\frac{10}{100}$ & $ 0.05245-0.002901 i $ & $ 0.002748-0.05837 i $ & $ 0.5967-0.1797 i $ \\
    \hline
         $\frac{11}{100}$ & $ 0.05418-0.002343 i $ & $ -0.01753-0.05885 i $ & $ 0.5380-0.2144 i $ \\
    \hline
         $\frac{12}{100}$ & $ 0.05581-0.001873 i $ & $ -0.03648-0.05863 i $ & $ 0.4846-0.2397 i $ \\
    \hline
         $\frac{13}{100}$ & $ 0.05735-0.001479 i $ & $ -0.05419-0.05783 i $ & $ 0.4365-0.2568 i $ \\
    \hline
         $\frac{14}{100}$ & $ 0.05881-0.001150 i $ & $ -0.07074-0.05657 i $ & $ 0.3935-0.2667 i $ \\
    \hline
         $\frac{15}{100}$ & $ 0.06020-0.0008767 i $ & $ -0.08619-0.05495 i $ & $ 0.3556-0.2703 i $ \\
    \hline
         $\frac{16}{100}$ & $ 0.06151-0.0006529 i $ & $ -0.1006-0.05305 i $ & $ 0.3228-0.2684 i $ \\
    \hline
         $\frac{17}{100}$ & $ 0.06276-0.0004721 i $ & $ -0.1140-0.05095 i $ & $ 0.2951-0.2616 i $ \\
    \hline
         $\frac{18}{100}$ & $ 0.06395-0.0003285 i $ & $ -0.1266-0.04872 i $ & $ 0.2725-0.2507 i $ \\
    \hline
         $\frac{19}{100}$ & $ 0.06508-0.0002173 i $ & $ -0.1382-0.04643 i $ & $ 0.2554-0.2362 i $ \\
    \hline
         $\frac{20}{100}$ & $ 0.06615-0.0001341 i $ & $ -0.1490-0.04415 i $ & $ 0.2438-0.2187 i $ \\
    \hline
    
     \end{tabular}
    \caption{Dimensionless form factor $1000\cdot\hat{B}$ of order $\alpha_s^2$ for the operator $O_1$, given at the 23 points in Eq. (\ref{eq:points}).
      The three columns are the coefficients of $\epsilon^{-2}$, $\epsilon^{-1}$ and $\epsilon^0$, respectively. The value of the $\epsilon^{-3}$ coefficient is
      $z$-independent and is equal to $-0.01012$ for all points.}
    \label{tab:table_O1}
\end{table}
\begin{table}[]
    \centering
    \begin{tabular}{|c|c|c|c|}
    \hline
     $O_2$ &  $\frac{1}{\epsilon^{2}}$ &  $\frac{1}{\epsilon^{1}}$ &  $\epsilon^{0}$  \\ [1ex] 
    \hline\hline
         $\frac{1}{1000}$ & $ 0.08632+0.1532 i $ & $ 0.05402+0.2669 i $ & $ 0.9238-4.765 i $ \\
    \hline
         $\frac{1}{500}$ & $ 0.08334+0.1455 i $ & $ 0.03352+0.2671 i $ & $ 0.3961-4.204 i $ \\
    \hline
         $\frac{1}{200}$ & $ 0.07466+0.1292 i $ & $ 0.02862+0.2737 i $ & $ -0.1371-3.136 i $ \\
    \hline
         $\frac{1}{100}$ & $ 0.06152+0.1107 i $ & $ 0.06222+0.2850 i $ & $ -0.2863-2.072 i $ \\
    \hline
         $\frac{2}{100}$ & $ 0.03880+0.08620 i $ & $ 0.1548+0.2984 i $ & $ -0.1014-0.8485 i $ \\
    \hline
         $\frac{3}{100}$ & $ 0.01922+0.06936 i $ & $ 0.2485+0.3023 i $ & $ 0.1907-0.1181 i $ \\
    \hline
         $\frac{4}{100}$ & $ 0.001802+0.05664 i $ & $ 0.3370+0.2998 i $ & $ 0.4826+0.3712 i $ \\
    \hline
         $\frac{5}{100}$ & $ -0.01399+0.04659 i $ & $ 0.4195+0.2924 i $ & $ 0.7541+0.7143 i $ \\
    \hline
         $\frac{6}{100}$ & $ -0.02848+0.03843 i $ & $ 0.4960+0.2815 i $ & $ 1.002+0.9578 i $ \\
    \hline
         $\frac{7}{100}$ & $ -0.04190+0.03170 i $ & $ 0.5669+0.2679 i $ & $ 1.225+1.129 i $ \\
    \hline
         $\frac{8}{100}$ & $ -0.05439+0.02609 i $ & $ 0.6325+0.2520 i $ & $ 1.425+1.244 i $ \\
    \hline
         $\frac{9}{100}$ & $ -0.06609+0.02138 i $ & $ 0.6933+0.2345 i $ & $ 1.604+1.314 i $ \\
    \hline
         $\frac{10}{100}$ & $ -0.07707+0.01741 i $ & $ 0.7494+0.2158 i $ & $ 1.761+1.348 i $ \\
    \hline
         $\frac{11}{100}$ & $ -0.08743+0.01406 i $ & $ 0.8012+0.1960 i $ & $ 1.899+1.351 i $ \\
    \hline
         $\frac{12}{100}$ & $ -0.09721+0.01124 i $ & $ 0.8489+0.1757 i $ & $ 2.017+1.328 i $ \\
    \hline
         $\frac{13}{100}$ & $ -0.1065+0.008874 i $ & $ 0.8927+0.1549 i $ & $ 2.115+1.284 i $ \\
    \hline
         $\frac{14}{100}$ & $ -0.1152+0.006897 i $ & $ 0.9328+0.1340 i $ & $ 2.195+1.220 i $ \\
    \hline
         $\frac{15}{100}$ & $ -0.1235+0.005260 i $ & $ 0.9694+0.1132 i $ & $ 2.256+1.139 i $ \\
    \hline
         $\frac{16}{100}$ & $ -0.1314+0.003918 i $ & $ 1.003+0.09278 i $ & $ 2.297+1.044 i $ \\
    \hline
         $\frac{17}{100}$ & $ -0.1389+0.002832 i $ & $ 1.033+0.07284 i $ & $ 2.318+0.9377 i $ \\
    \hline
         $\frac{18}{100}$ & $ -0.1460+0.001971 i $ & $ 1.060+0.05365 i $ & $ 2.319+0.8212 i $ \\
    \hline
         $\frac{19}{100}$ & $ -0.1528+0.001304 i $ & $ 1.084+0.03541 i $ & $ 2.299+0.6968 i $ \\
    \hline
         $\frac{20}{100}$ & $ -0.1592+0.0008049 i $ & $ 1.105+0.01837 i $ & $ 2.256+0.5670 i $ \\
    \hline

    \end{tabular}
    \caption{Dimensionless form factor  $1000\cdot\hat{B}$ of order $\alpha_s^2$ for the operator $O_2$, given at the 23 points in Eq. (\ref{eq:points}).
      The three columns are the coefficients of $\epsilon^{-2}$, $\epsilon^{-1}$ and $\epsilon^0$, respectively. The value of the $\epsilon^{-3}$
      coefficient is $z$-independent and is equal to $0.06073$ for all points.}
    \label{tab:table_O2}
\end{table}
We also present the values for $z \to 0$ separately, as the value at this point is not obtained the same way as for the 23 points,
but rather from the limit of the asymptotic expansion. We obtain
\begin{equation}
    1000 \cdot \hat{B}_{O_1}(0) = -\frac{0.01012}{\epsilon^3}+\frac{0.02486-0.02765 i}{\epsilon^2}+\frac{0.2802+0.09776 i}{\epsilon}+(0.6538+2.6231 i)
\end{equation}
\begin{equation}
    1000 \cdot \hat{B}_{O_2}(0) = \frac{0.06073}{\epsilon^3}+\frac{0.08846+0.1659 i}{\epsilon^2}+\frac{0.2020+0.2813 i}{\epsilon}+(4.3882-5.8342 i)
\end{equation}
In Figures \ref{fig:res_o1} and \ref{fig:res_o2} 
we present the $\epsilon^0$ coefficient's real and imaginary parts of expansions for $O_1$ and $O_2$, respectively. Both figures show
the ``switched expansions'' (with switching at $z=0.04$), as well as the values at the 23 points.
The bare form factors $\hat{B}$ associated with the operators $O_1$ and $O_2$
have recently been presented in  \cite{Fael:2023}\footnote{They are denoted there by $t_2^{Q_1}$ and $t_2^{Q_2}$, respectively.}.
These authors also worked out several expansions; they explicitly gave the result for the asymptotic expansion around $z=0$,
which is in perfect agreement with our result. The main difference between our work and ref. \cite{Fael:2023} lies in the methods used: While
they apply their ``expand and match'' approach as developed in \cite{Fael:2021kyg,Fael:2022rgm,Fael:2022miw} to construct the series expansions of the MIs,
we directly utilize \texttt{DESolver} and \texttt{DiffExp} to generate the expansions around $z=0$ and $z=1/10$.

\section{Summary}
\label{sec:conclusions}
\setcounter{equation}{0}

In this paper we worked out three-loop diagrams (of order $\alpha_s^2$) contributing
to the decay amplitude for $b \to s \gamma$ associated with the current-current operators
$O_1$ and $O_2$ at different values of $m_c$. As a continuation of the work done in \cite{Greub:2023msv}, we have calculated diagrams where at least
one gluon is touching the $b$-quark line (see \Fig{fig:diags_bleg} and \Fig{fig:diags_mixed}) in the present paper. We have also worked out all
three-loop diagrams with bubbles on the gluon lines (see \Fig{fig:diags_bubble}).
We have used \texttt{AMFlow} to calculate the boundary conditions for all diagrams at $z=1/100$,
as described in Section~\ref{sec:BoundaryConditions}. Using these boundary conditions, we were able to calculate the results at 23 different
points between $z=1/1000$ and $z=1/5$ with very high precision. We have also calculated the asymptotic expansion
around $z=0$ for all diagrams, which are useful in many practical applications. For the ``$b$-leg diagrams'',
the asymptotic expansion breaks down at $z \sim 0.07$, therefore we also worked out a Taylor expansion around
$z=1/10$, which corresponds to the physical value. For most diagrams the same expressions have also been calculated using \texttt{DiffExp}, and the results
using these two programs agree with very high precision. Our asymptotic expansions have been compared with the asymptotic expansions in \cite{Fael:2023},
and we noticed very good agreement.

The numerical results at the 23 points in Eq. (\ref{eq:points}) are given for $O_1$ and $O_2$
in Tables \ref{tab:table_O1} and \ref{tab:table_O2}, respectively.
These results and the symbolic expression for the asymptotic formula up to $z^{20}$ (as well as the symbolic expression for the Taylor expansion for the ``$b$-leg diagrams''
up to $(z-1/10)^{20}$) are given in electronic form in the file \texttt{ancillary.m}
which is submitted together with this paper. 

\section*{Acknowledgements}
\noindent
We  thank Daniel Wyler for general discussions related to radiative $B$-meson decays and for useful comments concerning the manuscript.
We would like to thank Martijn Hidding for very useful discussions concerning
his excellent program \texttt{DiffExp}. Many thanks also go to Xiao Liu for
answering our questions on his superb program \texttt{AMFlow} always in a very short time.
C.G. would like to thank Matteo Fael for very useful discussions related to his work \cite{Fael:2023}. 
Discussions with Valentin Hirschi, Nicolas Schalch and Yannick Ulrich on various features connected with multiloop diagrams are
also gratefully acknowledged.

\noindent
This work is partially supported by the Swiss National Science Foundation under grant 200020-204075.
H.M.A. and H.H.A. are supported by the Higher Education and Science Committee of Armenia 
Program Grant No.
21AG‐1C084.

\newpage

\appendix
\renewcommand{\theequation}{\Alph{section}.\arabic{equation}} 

\setcounter{equation}{0}

\section{Details on the ancillary file}
\subsection{Results for the individual contributions of different diagram classes to the form factor \texorpdfstring{$\hat{B}$}{B} in electronic form}
\label{app:Bi}

In the mathematica file ``ancillary.m'' (which is included in the submission of this paper) we give the contributions to the form factor
$\hat{B}$ as defined in Eq. (\ref{eq:Bhat}) for the 4 different classes of diagrams (``$b$-leg diagrams'', ``$s$-leg diagrams'', ``mixed diagrams'' and ``bubble diagrams''). 

All the results mentioned in the paper are given in this file. Firstly, the values at the 23 points in Eq. (\ref{eq:points}) are presented for all 4 classes of diagrams for both $O_1$ and $O_2$.
For the $O_1$ contributions, these are the expressions \texttt{BlegPointsO1}, \texttt{SlegPointsO1}, \texttt{MixedPointsO1} and \texttt{BubblesPointsO1} for the 4 classes of diagrams,
respectively (the same expressions for $O_2$ are named accordingly). The asymptotic expansions are also presented up to order $z^{20}$ for all classes of diagrams.
The corresponding expressions for $O_1$ are \texttt{BlegAsymO1}, \texttt{SlegAsymO1}, \texttt{MixedAsymO1} and \texttt{BubblesAsymO1}. As mentioned in the paper,
the Taylor expansion around $z=1/10$ is also presented for the ``$b$-leg diagrams'', as a function of $u$, where $u=z-1/10$, up to order $u^{20}$. This expression for $O_1$ is \texttt{BlegTaylorO1}.

Note that these formulas contain the following symbolic constants: $nl=3$ (number of `light' quarks), $nb=1$ (number of quarks with mass equal to $m_b$),
$nc=1$ (number of quarks with mass equal to $m_c$), $ca=3, cf=4/3, tr=1/2$ (color factors), $Qd=-1/3$ (charge of down-type quark) and $Qu=2/3$ (charge of up-type quark).
All these constants are given in the ancillary file.

At the end of the ancillary file, we present the two functions \texttt{FormFactorBhatO1} and 
\texttt{FormFactorBhatO2}, which, for a given value of $z$, return the form factor $\hat{B}$ corresponding to all 4 classes of diagrams, with all the symbolic constants inserted,
for $O_1$ and $O_2$, respectively. These functions automatically choose the correct expansion for the ``$b$-leg diagrams'' depending on whether the given value of $z$ is larger than $z=0.04$.


\end{document}